%% file: main.tex
\documentclass{vgtc}                          

\ifpdf
  \pdfoutput=1\relax                   
  \usepackage{graphicx}                
  \DeclareGraphicsExtensions{.pdf,.png,.jpg,.jpeg} 
\else
  \usepackage{graphicx}                
  \DeclareGraphicsExtensions{.eps}     
\fi%

\graphicspath{{figures/}{pictures/}{images/}{./}} 

\newcommand{\etal}{et~al.~} 
\newcommand{\ie}{i.e.,~}
\newcommand{\eg}{e.g.,~}
\newcommand{\ncorpus}{220 }

\usepackage{xurl}

\usepackage{subcaption}

\usepackage{stfloats}

\usepackage{graphicx}
\usepackage{wrapfig}
\usepackage{graphbox}
\usepackage[T1]{fontenc}
\usepackage{amsmath}  
\usepackage{multirow}
\usepackage{tabularx}

\usepackage[svgnames,x11names,table]{xcolor}
\usepackage{soul}
\usepackage{adjustbox}
\usepackage{hyperref}

\definecolor{c1}{HTML}{9796bb}
\definecolor{c2}{HTML}{00beb9}
\definecolor{c3}{HTML}{dfb0c7}
\definecolor{c4}{HTML}{add9a1}
\definecolor{c5}{HTML}{eac793}

\usepackage{microtype}                 
\PassOptionsToPackage{warn}{textcomp}  
\usepackage{textcomp}                  
\usepackage{mathptmx}                  
\usepackage{times}                     
\usepackage{cite}                      
\usepackage{tabu}                      
\usepackage{booktabs}                  

\onlineid{6460}

\vgtccategory{Research}

\vgtcinsertpkg

\title{More Than Beautiful: Exploring Design Features, Practical Perspectives, and Implications of Artistic Data Visualization}

\author{Xingyu Lan\thanks{Xingyu Lan is the corresponding author. She is a member of the Research Group of Computational and AI Communication at Institute for Global Communications and Integrated Media. e-mail: xingyulan96@gmail.com.}\\ %
        \scriptsize Fudan University %
\and Yifan Wang\thanks{e-mail: wangyifanlea@gmail.com}\\ %
     \scriptsize Fudan University %
\and Lingyu Peng\thanks{e-mail: lingyupeng6@163.com}\\ %
     \scriptsize \centering Harbin Institute of Technology
\and Xiaofan Ma\thanks{e-mail: xiaofanma\_13@163.com}\\ %
     \scriptsize \centering Sun Yat-sen University
     }

\abstract{Standing at the intersection of science and art, artistic data visualization has gained popularity in recent years and emerged as a significant domain. Despite more than a decade since the field's conceptualization, a noticeable gap remains in research concerning the design features of artistic data visualizations, the aesthetic goals they pursue, and their potential to inspire our community. To address these gaps, we analyzed \ncorpus data artworks to understand their design paradigms and intents, and construct a design taxonomy to characterize their design techniques (\eg sensation, interaction, narrative, physicality). We also conducted in-depth interviews with twelve data artists to explore their practical perspectives, such as their understanding of artistic data visualization and the challenges they encounter. In brief, we found that artistic data visualization is deeply rooted in art discourse, with its own distinctive characteristics in both inner pursuits and outer presentations. Based on our research, we outline seven prospective paths for future work.}

\CCScatlist{
  \CCScatTwelve{Human-centered computing}{Visualization design and evaluation methods}{}{}
}
\keywords{Artistic Visualization, Data Art, Visualization Design}

\begin{document}


\input{Sections/01-intro.tex}
\input{Sections/02-related.tex}
\input{Sections/03-case}

\input{Sections/04-survey}

\input{Sections/05-interview}

\input{Sections/06-discussion}

\input{Sections/07-conclusion}

\acknowledgments{
This work was supported by NSFC 62402121, Shanghai Chenguang Program, and Research and Innovation Projects from the School of Journalism at Fudan University.}

\bibliographystyle{abbrv-doi-hyperref-narrow}

\bibliography{template}
\end{document}

%% file: Sections/01-intro.tex
\firstsection{Introduction}
\maketitle

As a discipline that deals with visuals, data visualization has a natural affinity with art. Fueled by the rise of big data, artists are increasingly using digital tools and programming software to create data-driven artworks. In 2007, Viégas and Wattenberg~\cite{viegas2007artistic} proposed \textit{artistic data visualization} and defined it as ``visualizations of data done by artists with the intent of making art'', viewing it as an enchanting domain beyond visual analytics.
Today, as noted by a BBC article~\cite{bbc}, artists are ``making waves with big data'', and some data artworks have had a significant impact on society and the market (\eg the non-fungible token (NFT) series based on the artworks discussed in \autoref{ssection:machine} was sold for \$5.1 million at auction~\cite{yahoo}).
As an artist who has created a series of influential data artworks, Frankel~\cite{frankel1998envisioning} once wrote in \textit{Science} that she realized artworks can achieve significant consequences, especially ``to communicate important information about science research not only to other scientists in the lab, or in the field, but to a broader, nonscientific public, as well.'' 

However, compared to its flourish in the industry, research focused on artistic data visualization remains relatively scarce. As argued by Kosara~\cite{kosara2007visualization}, the ``classifications of visualization are often based on technical criteria, and leave out artistic ways of visualizing information''. 
To the best of our knowledge, formative research in artistic data visualization (\eg full papers published in high-quality journals or conferences) is not very abundant, with most studies focusing on specific applications, such as addressing issues like energy use and environmental pollution~\cite{rodgers2011exploring,schroeder2015visualization,perovich2020chemicals}. 
By contrast, theoretical research on artistic data visualization appears insufficient, leaving many questions underexplored. These include: What are the design features of artistic data visualizations? What values and viewpoints do data artists hold? How might artistic data visualization inspire or benefit the visualization community?

To address these gaps, this work undertook research in three main steps. First, we collected a corpus of artistic data visualizations to identify representative cases (\autoref{sec:case}) and characterize their design features by analyzing their design paradigms, intents, and techniques (\autoref{sec:space}). Next, we interviewed twelve data artists to gain deeper insights into their practical experiences and perspectives on artistic data visualization, such as how they describe their aesthetic pursuits and respond to potential critiques (\autoref{sec:interview}).
Based on these findings, we discuss the implications of artistic data visualization for our community and introduce future research opportunities (\autoref{sec:discuss}). The corpus can be visually browsed at \url{https://artisticvis.github.io/}, and the raw datasets and codes can be accessed at \url{https://bit.ly/47ER8aJ}.

This work takes an initial step toward understanding artistic data visualization, an important form that expands visualization applications and blends technology with art. In summary, we contribute (i) empirical understanding about the design paradigms, intents, and existing techniques of artistic data visualizations by analyzing a corpus of artworks, (ii) first-hand insights from practitioners that provide valuable perspectives and foster community dialogue, and (iii) guidance for future work to explore new possibilities and advancements, inspired by the dialogue between art and science.

%% file: Sections/02-related.tex
\section{Background and related work}

This section reviews the background on artistic data visualization and previous research related to this topic.

\subsection{Artistic Data Visualization in Art History Context}
\label{ssec:contemporary}

Art history has been marked by transformative periods characterized by different aesthetic pursuits, materials, and methods. Since the time of Plato, imitation (or \textit{mimesis}, which views art as a mirror to the world around us) has been an important pursuit~\cite{pooke2021art}. Successful artworks, such as Greek sculptures and the convincing illusions of depth and space in Renaissance paintings, exemplify this tradition.
The advent of modern society and new technology, especially photography, posed a significant challenge to the notion of art as imitation~\cite{perry2004themes}. By the 1850s, modern art began to emerge with the core goal of transcending traditional forms and conventions. Movements like Post Impressionism, Expressionism, and Cubism revolutionized art through innovative uses of form (\eg color, texture, composition), moving art towards abstraction and experimentation. 
After World War II, the Cold War and the surge of various social problems heightened skepticism about the progress narrative of modernity and the superiority of the capitalist system, leading to the rise of postmodernism and the birth of contemporary art~\cite{hopkins2000after,harrison1992art}. One prominent feature of contemporary art is the absence of fixed standards or genres historically defined by the church or the academy. Postmodern design neither defines a cohesive set of aesthetic values nor restricts the range of media used~\cite{pooke2021art}, sparking novel practices such as installations, performances, lens-based media, videos, and land-based art~\cite{hopkins2000after}.
Meanwhile, artists have increasingly drawn energy from various philosophical and critical theories such as gender studies, psychoanalysis, Marxism, and post-structuralism~\cite{pooke2021art}. As a result, as described by Foster~\cite{foster1999recodings}, artists have increasingly become ``manipulators of signs and symbols... and the viewer an active reader of messages rather than a passive contemplator of the aesthetic''. Hopkins~\cite{hopkins2000after} described this shift as the ``death of the object'' and ``the move to conceptualism''. 

Emerging within the contemporary art historical context, data art has been significantly propelled by the advent of big data. An early milestone was Kynaston McShine's 1970 exhibition \textit{Information} at the Museum of Modern Art (MoMA). 
In 2008, Google’s Data Arts Team was founded to explore the advancement of what creativity and technology can do together~\cite{google}.
In 2012, Viégas and Wattenberg's \textit{Wind Map}, an artwork that visualizes real-time air movement, became the first web-based artwork to be included in MoMA's permanent collection~\cite{wind}.
Since 2013, the academic conference IEEE VIS has included an Arts Program (IEEE VISAP), showcasing artistic data visualizations through accepted papers and curated exhibitions. 
As noted by Barabási~\cite{dataism} (a Fellow of the American Physical Society and the head of a data art lab), data has become a vital medium for artists to deal with the complexities of our society: ``Humanity is facing a complexity explosion. We are confronted with too much data for any of us to make sense of...The traditional tools and mediums of art, be they canvas or chisel, are woefully inadequate for this task...today’s and tomorrow’s artists can embrace new tools and mediums that scale to the challenge, ensuring that their practice can continue to reflect our changing epistemology.''


\subsection{Research on Artistic Data Visualization}
\label{ssec:artisticvis}

Artistic data visualization is also referred to as artistic visualization, data art, or information art~\cite{holmquist2003informative,rodgers2011exploring,few,viegas2007artistic}. Despite the varying terminologies, there is a basic consensus that artistic data visualization must be art practice grounded in real data~\cite{viegas2007artistic}.
Since the early 2000s, a series of papers introduced innovative artistic systems for visualizing everyday data, such as museum visit routes and bus schedule information~\cite{skog2003between,holmquist2003informative,viegas2004artifacts}.
In 2007, Viégas and Wattenberg~\cite{viegas2007artistic} explicitly proposed the concept of \textit{artistic data visualization} and viewed it as a promising domain beyond visual analytics.
Kosara~\cite{kosara2007visualization} drew a spectrum of visualization design, positioning artistic visualization and pragmatic visualization at opposite ends of this spectrum to demonstrate that the goals of these two types of design often diverge. 
However, overall, research on this subject has been sparse. Among those relevant papers, most have focused on specific applications of artistic data visualization. 
For instance, Rodgers and Bartram~\cite{rodgers2011exploring} utilized ambient artistic data visualization to enhance residential energy use feedback. Samsel~\etal~\cite{samsel2018art} transferred artistic styles from paintings into scientific visualization.
Artistic practice has also been observed in fields such as data physicalization~\cite{hornecker2023design,perovich2020chemicals,offenhuber2019data} and sonification~\cite{enge2024open}. For example, Hornecker~\etal~\cite{hornecker2023design} found that many artists are practicing transforming data into tangible artifacts or installations. Enge~\etal~\cite{enge2024open} discussed a set of representative artistic cases that combine sonification and visualization.


On the other hand, some studies, while not focusing on artistic data visualization, have explored a key art-related concept: aesthetics. 
For example, Moere~\etal~\cite{moere2012evaluating} compared analytical, magazine, and artistic visualization styles, noting that analytical styles enhance the discovery of analytical insights, while the other two induce more meaning-based insights. Cawthon~\etal~\cite{cawthon2007effect} asked participants to rank eleven visualization types on an aesthetic scale from ``ugly'' to ``beautiful'', finding that some visualizations (\eg sunburst) were perceived as more beautiful than others (\eg beam trees).
Factors such as embellishment~\cite{bateman2010useful}, colorfulness~\cite{harrison2015infographic}, and interaction~\cite{stoll2024investigating} have also been found to influence aesthetics. 
However, most of these studies have simplified aesthetics to hedonic features (\eg beauty), without delving into the nuanced connotations of aesthetics.

The value of artistic data visualization to the visualization community is still in controversy. For instance, Few~\cite{few} argued for a clearer distinction between data art and data visualization; he highlighted the negative consequences when data art ``masquerades as data visualization'', such as making visualizations difficult to perceive. Willers~\cite{willers2014show} thought the artistic approach is ``unlikely be appreciated if the intention was for rapid decision making.''
To address these gaps, more empirical research needs to be conducted to explore how artistic data visualizations are designed, their underlying pursuits, and how they might inspire our community.

%% file: Sections/03-case.tex
\section{Case Examples}
\label{sec:case}



As artistic data visualization is a relatively under-explored field in academia, we follow prior work~\cite{segel2010narrative,shi2021communicating} and first introduce three typical cases to provide an initial exploration and motivation for the study of artistic data visualization. This approach not only offers a preliminary understanding of the practices in this domain, but also provides representative examples from the corpus analyzed in \autoref{sec:space}, helping to contextualize the more abstract analysis presented later.
These cases were selected based on three main criteria: (i) we searched public materials about the artworks in \autoref{sec:space} and identified those that have been exhibited in top-tier exhibitions (\eg having a work exhibited at MoMA in New York is regarded as a high recognition for artists), (ii) works that have been widely reported or recognized by news media, (iii) artists that have received prestigious career awards. Based on these criteria, we selected three works that best meet them and span a considerable range of time.


\subsection{Case I: Particle Dreams in Spherical Harmonics}
\label{ssection:particle}

\begin{figure}[h]
 \centering
 \includegraphics[width=\columnwidth]{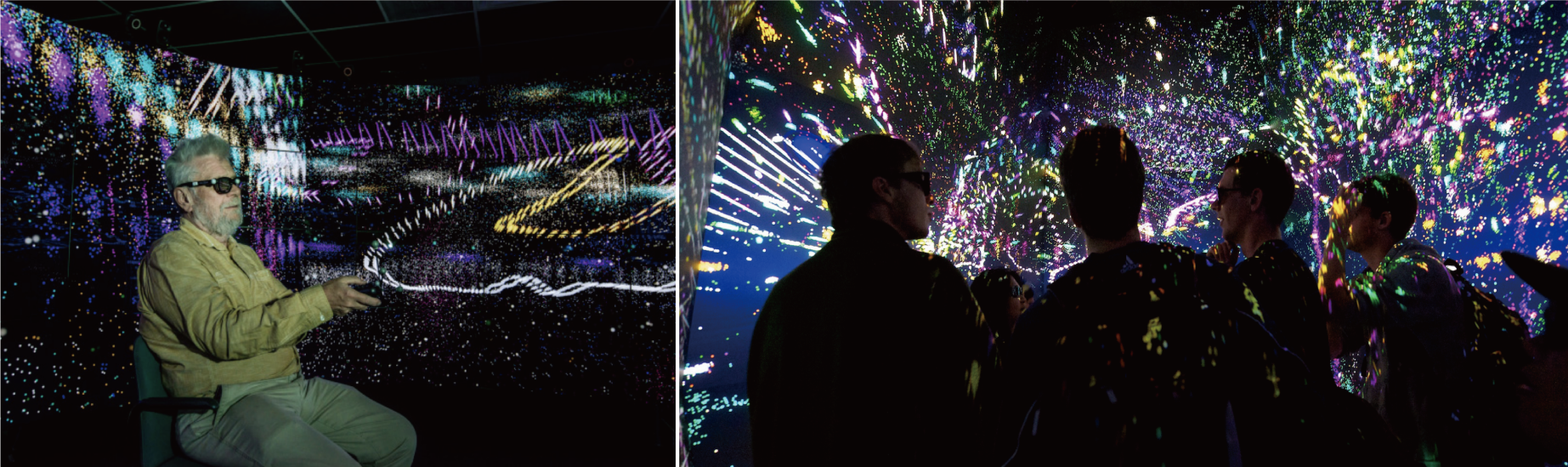}
 \vspace{-2em}
 \caption{Particle Dreams in Spherical Harmonics~\cite{sandin}.}
 \label{fig:cases_1}
 \vspace{-0.5em}
\end{figure}

Particle Dreams in Spherical Harmonics is a Virtual Reality (VR) artwork produced by Dan Sandin and his team. As an internationally recognized pioneer of electronic art and visualization~\cite{sandin}, Sandin is the director emeritus of the Electronic Visualization Laboratory (EVL) and a professor emeritus in the School of Art and Design at the University of Illinois Chicago. With a background in physics, Sandin has dedicated his career to exploring the intersection of technology and art.
In the 1970s, he developed the Sandin Image Processor, a highly programmable analog computer for processing live video feeds. It was one of the early devices that allowed artists to manipulate video data inputs in real-time, solving the problem of computer-graphics systems being too expensive and not easily accessible to most people~\cite{johnson2024electronic}.
Since the 1990s, Sandin and his colleagues began developing the CAVE (Cave Automatic Virtual Environment), a pioneering VR theater system that provides immersive experiences in a 3D space where computer-generated imagery is projected onto walls and floors.


Sandin himself also used these technologies to produce a series of data artworks, particularly in the realm of scientific visualization. \autoref{fig:cases_1}, for example, is a VR artwork based on the ``physical simulation of over one million particles with momentum and elastic reflection in an environment with gravity.''~\cite{sandin}
It also creates a multisensory experience by generating sound that is triggered and modified by the user-particle interactions in real-time.
A viewer commented that standing in the artwork ``was like standing in a rainstorm made of rainbow fragments, with the power to guide the storm by hand. It was unsettling, out-of-body, very trippy stuff, a powerful artistic experience.''~\cite{evl_vr}
When talking about the motivations of their work, Sandin put more emphasis on building a ``more effective communication medium and a much more effective way to visualize data''~\cite{evl_synthesis}.
This is also the philosophy of his lab, that is, ``systems should be user-oriented (easy to use, easy to learn), low-cost, interactive, and real-time (to provide immediate feedback).''~\cite{johnson2024electronic}
As for himself as an artist, Sandin adopted a rather open attitude to creativity: ``About creativity—my personal view of it is kind of like I’m a pipe or conduit. And all this stuff just happens to be flowing through me because I’ve chosen to position myself in that flow. I have no problem with the word `creation' as long as people don’t lay too much molasses on it.''~\cite{vdb}

\subsection{Case II: Smell Maps}
\label{ssection:smell}

\begin{figure}[h]
 \centering
 \includegraphics[width=\columnwidth]{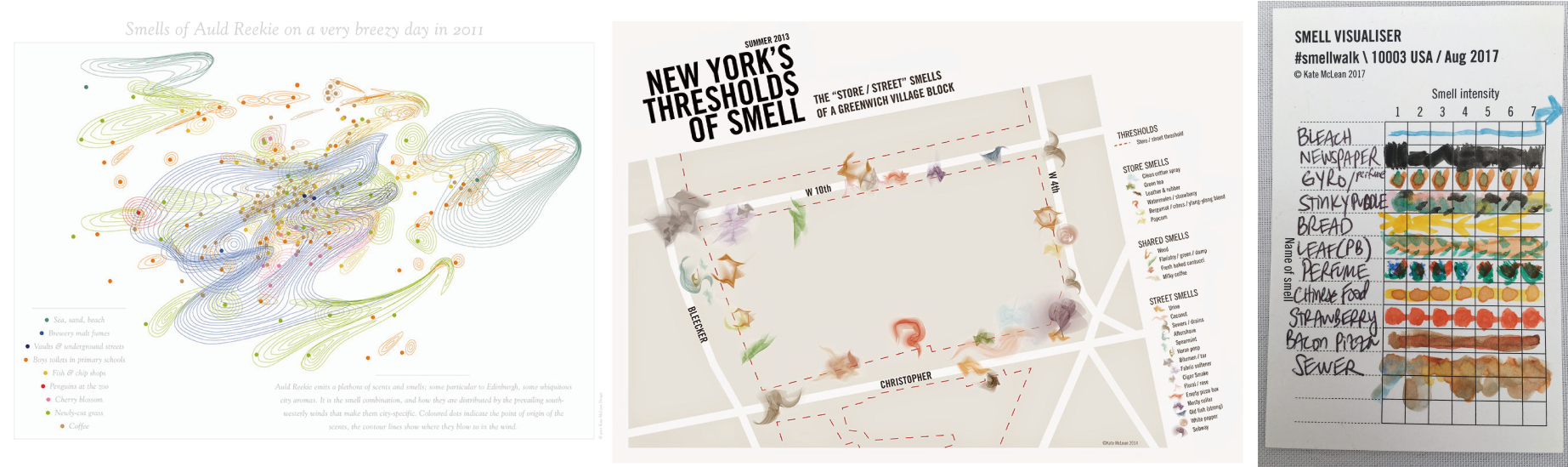}
 \vspace{-2em}
 \caption{Smell Maps~\cite{smellmaps}.}
 \label{fig:cases_2}
 \vspace{-1em}
\end{figure}

Since 2010, British artist Kate McLean has been working on translating the sensed aspects of place into visualizations. Starting with the first smell map of Paris, she has created a set of smell maps in various cities (\eg in \autoref{fig:cases_2}, the first two maps show Edinburgh and New York, respectively).
She used olfactory walks to collect data by first selecting specific routes in the cities, and then recruiting volunteers for these walks. The characteristics of the smells (\eg name, intensity) were recorded by the volunteers using smell notes (\autoref{fig:cases_2}, right). She also designed activities such as ``smell catching'' (noticing distant, airborne smells when walking) and ``smell hunting'' (searching for the sources of smell) to spark participants' interest and sensational involvement during walks~\cite{mclean2019nose}. 

Smell maps were motivated by the urban research by Douglas Porteous and Charles Foster~\cite{mclean2019nose}, who offered profound reflections and criticisms on contemporary living spaces, such as our alienation from physical experiences and over-reliance on vision.
Porteous theorized the concept of the \textit{smellscape}, pointing out that ``like visual impressions, smells may be spatially ordered or place-related''. But unlike an ordered visual landscape, the smellscape is an emotive environment that is ``non-continuous, fragmentary in space and episodic in time and limited by the height of our noses from the ground, where smells tend to linger.''~\cite{porteous1985smellscape}
Influenced by these theories, McLean's smell maps have placed a strong emphasis on physical participation and personal interpretation. She based all her work ``on physical experiences, rather than algorithms,''~\cite{cnn} and visualized these individual, subtle, and subjective olfactory data through artistic expressions (\eg using colored spots and rippling circles to present the sources and diffusion of smells). 
When designing visuals, McLean did not seek singular, precise scientific results but rather aimed for ``a visual synthesis of the different experiences reported by smellwalkers...I am interested in negotiating different perceptions.''~\cite{atlas}
In essence, she considered her practice and research to be qualitative~\cite{marie}: ``My aim is to celebrate and highlight the subjective elements of human perceptions of the smellscape...I never claim that the scents are objective, and my research to date indicates that it may not even be possible.''


\subsection{Case III: Machine Hallucination}
\label{ssection:machine}

\begin{figure}[h]
 \centering
 \includegraphics[width=\columnwidth]{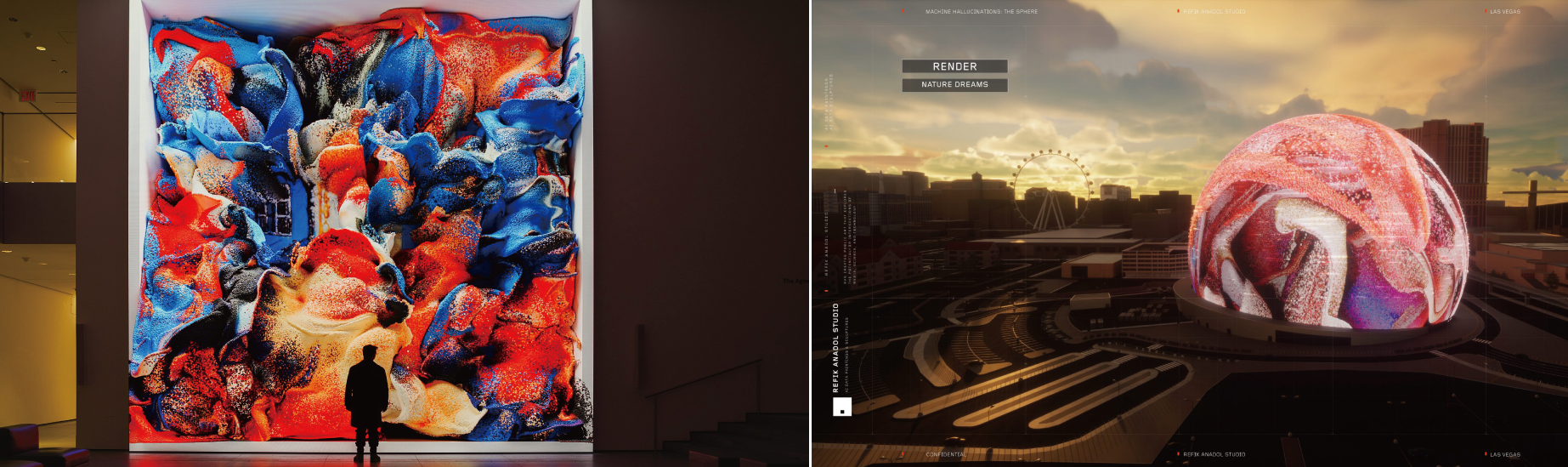}
 \vspace{-2em}
 \caption{Machine Hallucinations~\cite{machine}.}
 \label{fig:cases_3}
 \vspace{-2em}
\end{figure}

Machine Hallucination is a multi-series art project by Refik Anadol and his studio, utilizing big data and artificial intelligence (AI) to create immersive art experiences.
For example, \textit{Machine Hallucination - NYC}, as the first artwork in this series, employed machine-learning algorithms to process over 100 million photographs of New York. \textit{Machine Hallucination - Coral Dreams} was based on the training of more than 35 million images of coral. \textit{Machine Hallucination - Unsupervised} explored MoMA's vast collection, encompassing 150 years and nearly 140,000 art pieces. It was transformed into what Anadol called a ``living data sculpture'': a piece of artwork that is constantly changing, projecting an infinite number of alternative artworks that the machine creates in real-time across a giant media wall~\cite{yahoo} (\autoref{fig:cases_3}, left).
In a 2021 interview~\cite{momainterview}, Anadol said, ``In the past five years, we’ve trained more than 100 AI models, and used close to five petabytes of raw data. This is, as far as I know, one of the most challenging datasets ever used beyond specific research contexts, from clouds to national parks to cities to urban studies of Seoul, Stockholm, Berlin, Istanbul, New York, Los Angeles.'' 
Characterized by blossoming colors, biomechanical shapes, and constantly evolving data patterns, these artworks have achieved immense success. They have been projected onto notable architectural landmarks, such as the Walt Disney Concert Hall, Casa Battló, and the 580,000-square-foot Las Vegas Sphere (\autoref{fig:cases_3}, right).

Anadol's artwork demonstrates a strong pursuit of machine aesthetics. In a TED talk~\cite{ted}, he described his motivation for creating such artworks: ``Can data become a pigment? This was the very first question we asked when starting our journey to embed media arts into architecture, to collide virtual and physical worlds. So we began to imagine what I would call the poetics of data.'' Specifically for Machine Hallucination, it is an ambitious experiment concerning whether machines can dream and an attempt to re-present how machines interpret vast amounts of data in its ``brain'' (``AI in this case is creating this pigment that doesn't dry, a pigment that is always in flux, always in change, and constantly evolving and creating new patterns.''~\cite{yahoo})
While his works are highly technical, Anadol believes: ``Artificial intelligence is a mirror for humanity...It’s all about who we are as humans.''~\cite{fastcompany2}



%% file: Sections/04-survey.tex
\section{Design Features of Artistic Data Visualization}
\label{sec:space}

In this section, we analyzed \ncorpus artworks to outline a bigger picture of artistic data visualizations' design features. 

\subsection{Corpus Collection}

\begin{figure*}[b]
 \centering
 \includegraphics[width=0.94\textwidth]{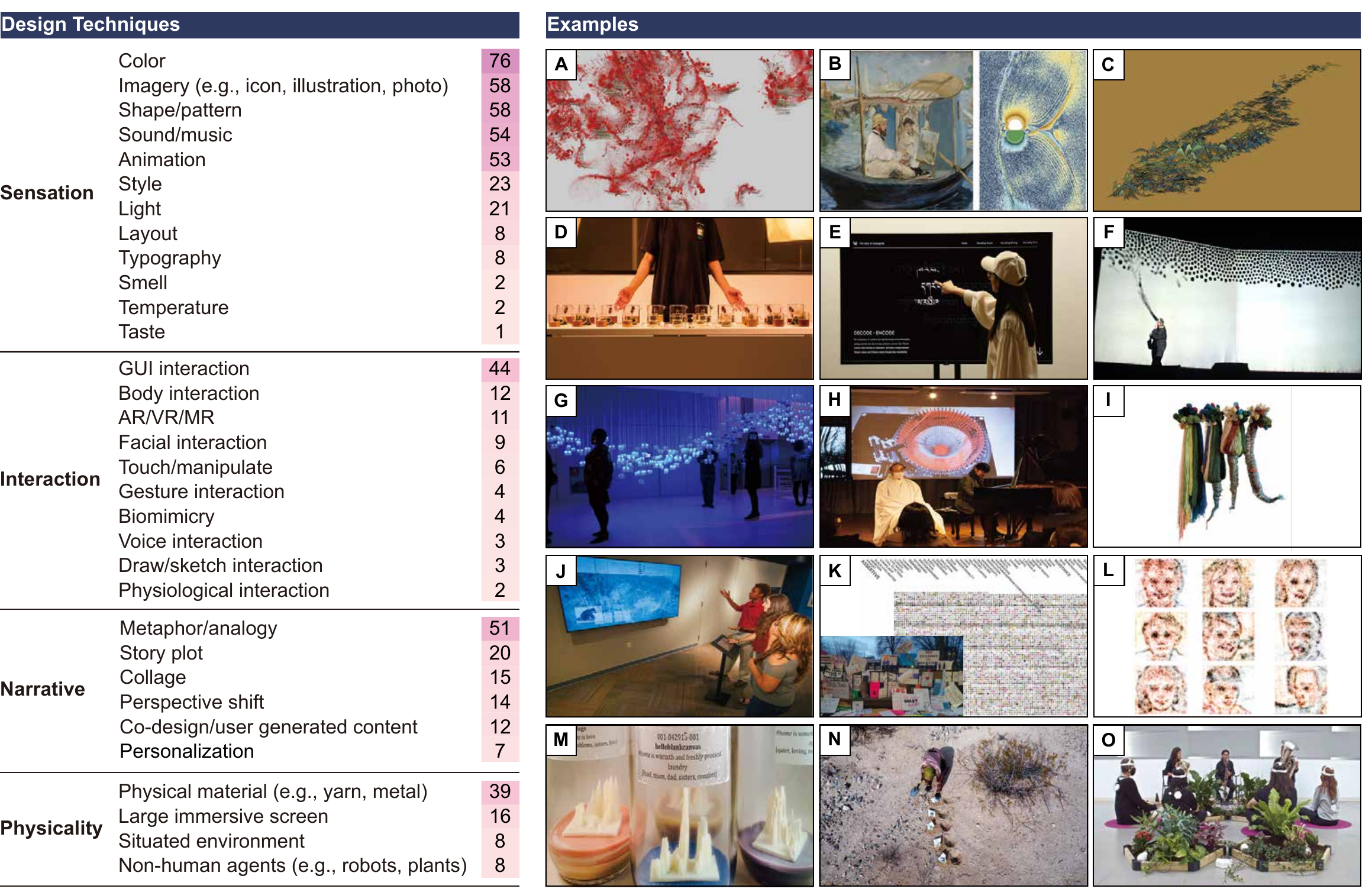}
 \caption{Left: All identified design techniques and their frequencies. Right: Examples of the artworks. (A) Agitato~\cite{agitato}, (B) Applying color palettes in paintings to scientific visualization~\cite{samsel2018art}, (C) Shan Shui in the World~\cite{shi2016shan}, (D) Bitter Data~\cite{li2023bitter}, (E) Decoding • Encoding~\cite{tibetan}, (F) Messa di Voce~\cite{messadivoce}, (G) Bion~\cite{bion}, (H) NeuroKnitting Beethoven~\cite{neuro}, (I) Climate Prisms~\cite{prisms}, (J) Oceanforestair~\cite{oceanforestair}, (K) Art of the March~\cite{protest}, (L) Decomposition of Human Portraits~\cite{face}, (M) \#home~\cite{home}, (N) The Sky is Falling~\cite{sky}, (O) Beyond Human Perception~\cite{plants}.}
 \label{fig:techniques}
 \vspace{-2em}

\end{figure*}

We began by collecting artworks from the IEEE VISAP (VIS Arts Program), which is a program associated with the top-tier visualization conference IEEE VIS. 
IEEE VISAP was started in 2013 and has run for more than ten years. 
Over the years, IEEE VISAP has become the leading event for artistic data visualization and has formed a representative collection of artworks.
We scraped all the accepted artworks of IEEE VISAP from its official website, resulting in a total of 231 non-duplicate artworks. After excluding 36 artworks that are now inaccessible on the web, we kept 195 artworks published between 2013 and 2023.
Overall, we believe that IEEE VISAP is a good source because: (i) It offers a publicly accessible dataset dedicated to artistic data visualization. (ii) Its review process ensures the artworks' relevance to data visualization. (iii) The inclusion of textual explanations or short papers by the artists provides essential materials for our analysis.
We also searched venues such as the Info+ Conference, Tapestry Conference, EyeO Festival, and Information is Beautiful. However, although these venues feature active data artists, they either lack a dedicated track explicitly for artistic data visualization or do not include designers' explanatory materials.
In the end, we opted to use a snowballing technique to identify literature that introduces the practices of artistic data visualization and to include as many data artworks as possible.
As a result, we added 25 artworks to our corpus, bringing the total to \ncorpus artworks. The additional artworks were mostly published in non-VIS tracks.

The publication time of these artworks spans from 1996 to 2023 (\includegraphics[align=b,scale=0.13]{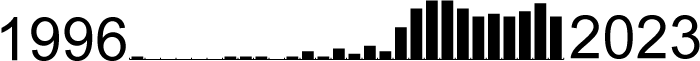}).
We in total identified 516 distinct authors of these artworks coming from 188 different affiliations, and their backgrounds are also diverse (see \autoref{tab:authors} for more details).
Each artwork has an average of 2.38 authors. Among the \ncorpus works, 85 have only one author. Of the remaining 135 collaborative works, 87 (64.44\%) were created by cross-disciplinary teams.
Another interesting finding is that apart from reporting their official occupations (\eg professor, PhD student), 115 authors used personalized labels to define themselves, such as ``multimedia artist and roboticist'', ``an intermedia artist and an acknowledged pioneer'', ``artist, technologist'', and ``musician''.
17 authors simultaneously hold academic and industry positions but place their industry identity ahead of their academic identity (\eg ``multidisciplinary digital media artist and university professor", "media artist and researcher'') showing that their role as practitioners is highly valued.

\begin{table}[h]
\centering
\vspace{-0.5em}
\caption{Affiliations and backgrounds of the authors from the corpus.}
\vspace{-1em}
\begin{subtable}[t]{0.56\linewidth} 
\fontsize{6.8pt}{7pt}\selectfont
\centering
\begin{tabular}{@{}p{0.8\linewidth}p{0.12\linewidth}@{}}
\toprule
\textbf{Affiliation (Top 5)}                     & \textbf{Num} \\
\midrule
University of California, Santa Barbara & 23 \\
Massachusetts Institute of Technology   & 21 \\
Northeastern University                 & 21 \\
University of Texas, Austin             & 19 \\
University of Calgary                   & 14 \\
\bottomrule
\end{tabular}
\end{subtable}%
\hfill
\begin{subtable}[t]{0.41\linewidth} 
\fontsize{6.5pt}{7pt}\selectfont
\centering
\begin{tabular}{@{}p{0.5\linewidth}p{0.4\linewidth}@{}}
\toprule
\textbf{Background}                     & \textbf{Num} \\

\midrule
Art/Design        & 193 (37.40\%) \\
Engineering        & 132 (25.58\%) \\
Cross-Disciplinary & 82 (15.89\%) \\
Other Disciplines              & 66 (12.79\%) \\
Unknown            & 43 (8.33\%) \\
\bottomrule
\end{tabular}
\end{subtable}
\vspace{-2em}
\label{tab:authors}
\end{table}







\subsection{Design Analysis}


\subsubsection{Analysis Method}

We analyzed the design features of the artworks using open coding while referring to the artists' own explanations.
Two authors were responsible for the coding process, focusing on two main aspects~\cite{shi2021communicating,sarikaya2018we}: (i) What are the design intents of the artworks? and (ii) How are they designed? First, we familiarized ourselves with the common naming and categorization of visualization design intents and techniques proposed in prior design taxonomies (\eg ~\cite{lan2023affective}).
Then, we went through the artworks independently and generated codes to describe their design intents and techniques. Similar codes were categorized into groups. For example, we identified multiple channels of designing interaction, such as \textit{GUI interaction}, \textit{facial interaction}, and \textit{voice interaction}. Thus, these codes were grouped into a category called \textit{interaction}.
Next, we met to compare codes, discuss mismatches, and adjust inappropriate codes. For example, we initially used multiple codes to describe various materials used in the artworks, such as \textit{plastic}, \textit{metal}, \textit{yarn}, and \textit{glass}. During our discussion, we found that these codes were too detailed and difficult to enumerate exhaustively. Therefore, we consolidated them into a single code called \textit{physical materials}. Additionally, we observed that some artists (though not all) used higher-level terms to describe the fundamental concepts or philosophies behind their artworks. We agreed that this information is valuable and complements low-level design techniques. As a result, we added a new dimension called \textit{design paradigm} to our codebook and proceeded with another round of coding. In other words, our final taxonomy contains three main dimensions: design paradigms, design intents, and design techniques. After four rounds of meetings and coding, we achieved 100\% agreement on the coding scheme.

\subsubsection{Design Paradigms}
\label{sssec:paradigm}
We identified 37 explicit mentions of high-level design paradigms. The most frequently mentioned paradigm is \textit{generative art} (N = 12), which refers to art created using autonomous systems or algorithms.
Other mentions include concepts such as \textit{abstract art} (2), \textit{physical computing} (2), \textit{algorithmic art} (2), \textit{computational aesthetics} (2), \textit{digital art} (2), and \textit{digital fabrication} (2).
For example, parametric design generates artworks through a series of parameters. Designers can flexibly adjust and control the parameters to explore a range of design variations. Glitch art is a paradigm that celebrates errors in computer programs and embraces the aesthetics of imperfections.
A common feature of the above paradigms is their emphasis on the role of computers.

We also identified another set of paradigms that are more human-centered, such as \textit{speculative art} (2), \textit{digital humanism} (2), \textit{participatory design} (1), \textit{aministic design} (1), and \textit{neo-concrete art} (1). For instance, speculative art explores alternative realities and futures, often through experimental practices within communities. Participatory design involves people actively in creating artwork to promote social inclusion, empower marginalized communities, and reflect their needs and values. 
When contextualized within art history~\cite{pooke2021art,hopkins2000after,perry2004themes}, nearly all the aforementioned paradigms fit the scope and trends of contemporary art.


\subsubsection{Design Intents}
\label{sssec:intents}

We initially coded the design intents of artistic data visualizations based on Lan~\etal~\cite{lan2023affective}'s work, which identified ten types of design intents (\eg \textit{inform}, \textit{engage}, \textit{experiment}) in affective visualization design.
As a result, although all the ten intents are present in our corpus, we also identified five new intents (marked using * in the following text).
In terms of frequency, the most common intent is \textit{experiment} (N = 49).
The core spirit of this intent is to challenge traditional notions of data visualization and explore unconventional or novel representations.
Other design intents include \textit{inform} (39), \textit{engage} (36), \textit{re-present*} (36), \textit{provoke} (30), \textit{criticize*} (19),  \textit{equip*} (13), \textit{analyze*} (12), \textit{advocate} (10), \textit{socialize} (7), \textit{witness*} (6), \textit{archive} (5), \textit{commemorate} (3), and \textit{empower} (3). 
For example, 
\autoref{fig:techniques} (A) \textit{re-presents} the experience of listening to music by transforming the artist's subtle emotions into animated metaphorical shapes. We use the term \textit{re-present} instead of \textit{represent} to highlight the artist's intent to capture and reveal something invisible while integrating their own interpretations.
\autoref{fig:techniques} (N) exemplifies the newly identified intent, \textit{witness}. To show the number of civilians killed by US drone strikes, the artist performed in a desert from dawn to dusk, commemorating each civilian with an earth mound, a white cloth, a stone, and a prayer. Viewers witnessed the ceremony through live streams.
As another example, the intent of \autoref{fig:techniques} (B) is mainly to \textit{equip} artists with a tool for coloring scientific visualizations using expressive palettes extracted from paintings. This also resonates with the case of Sandin (\autoref{ssection:particle}) and previous findings that artists sometimes write software themselves or participate in software development~\cite{li2021we}.



\subsubsection{Design Techniques}
\label{sssec:techniques}
A total of 32 design techniques were identified. The 32 design techniques were further grouped into four categories: \textit{sensation}, \textit{narrative}, \textit{interaction}, and \textit{physicality} (see \autoref{fig:techniques}).

\textbf{Sensation}. This category primarily focuses on creating various visual, auditory, olfactory, and other sensory effects. Relevant techniques include the use of \textit{color}, \textit{imagery}, \textit{shape/pattern}, \textit{sound/music}, \textit{animation}, \textit{style}, \textit{light}, \textit{layout}, \textit{typography}, \textit{temperature}, \textit{smell}, and \textit{taste}. 
For example, \autoref{fig:techniques} (A) transformed the evolving music listening experience into a generative artwork composed of vibrant colors, organic shapes, and animation.
\autoref{fig:techniques} (B) applied color palettes extracted from paintings to scientific visualizations.
\autoref{fig:techniques} (C) transformed the map of New York City into a style reminiscent of traditional Chinese painting.
\autoref{fig:techniques} (D)~\cite{li2023bitter} transforms 100,000 distress postings using data edibilization. The data was mapped to the bitterness and color of tea, enabling users to observe, smell, and taste the distress on social media.


\textbf{Interaction}. 
Among interactive techniques, GUI interaction, such as allowing users to click on or scroll a website, mobile app, or tablet, is most common (\eg \autoref{fig:techniques} (E)). 
Other relevant techniques include \textit{body interaction}, \textit{facial interaction}, \textit{touch/manipulate}, \textit{gesture interaction}, \textit{biomimicry}, 
\textit{AR/VR/MR}, \textit{voice interaction},  \textit{draw/sketch interaction}, and \textit{physiological interaction}. 
For example, \autoref{fig:techniques} (F) is a performance driven by voice interaction. The performers made various sounds and the sounds were transformed into the real-time animated visualization in the background. 
\autoref{fig:techniques} (G) is composed of hundreds of ``bions'' (individual elements of primordial biological energy) programmed according to biomimicry sensing. When a viewer approaches the installation, bions quickly communicate to each other, but eventually they become accustomed to the stranger's presence and respond as if he/she is part of their ecosystem.
\autoref{fig:techniques} (H) is a neuroknitting artwork that utilizes people's brainwave data when listening to music to drive the knitting machine.

\textbf{Narrative}. This category encompasses storytelling methods such as \textit{metaphor/analogy}, \textit{story plot}, \textit{collage}, \textit{perspective shift}, \textit{co-design/user generated content}, and \textit{personalization}. 
For example, 
\autoref{fig:techniques} (I) uses yarn to represent cumulative emissions over years; as the emission gets larger and larger, the yarn gets tighter and tighter, metaphorically mimicking a sense of suffocation.
\autoref{fig:techniques} (J) tells a data story about the shrinking of Arctic ponds by crafting a prepared storyline.
\autoref{fig:techniques} (K) is a website that collages more than 6,000 signs of the Boston Women’s March, which silently demonstrates the collective efforts of protesters. 
\autoref{fig:techniques} (L) employs a perspective shift to visualize human faces from an unconventional perspective—the eye of a deep neural network.
\autoref{fig:techniques} (M) invited visitors to share keywords about their homes. The keywords were used to filter a live Twitter stream, and the locations of these tweets were 3D printed as physical maps, which were personalized to each visitor.


\textbf{Physicality}. This category contains techniques that augment physical experiences, including \textit{physical materials} (\eg yarn, metal), \textit{large immersive screen}, \textit{situated environment}, 
\textit{non-human agents (\eg \textit{plants}, \textit{robots}, \textit{bacteria}).}
For example, the two aforementioned artworks, \autoref{fig:techniques} (I, M), use yarn and 3D printing to physicalize data, respectively.
\autoref{fig:techniques} (N), as introduced earlier, was an art performance conducted at a situated location, a desert, to provide viewers with an authentic sense of the environment affected by drone strikes.
Lastly, several artworks also utilize non-human agents, including entities that perform tasks or functions typically associated with humans. For example, \autoref{fig:techniques} (O), for example, played music to plants; sensor data from the plants was collected to explore the connection between technology and plant life.

\subsection{Observations}

Although the design of artistic data visualization shares some commonalities with other realms (\eg techniques like metaphor, story plot, personalization, and GUI interaction have also been found in narrative visualization~\cite{segel2010narrative,shi2021communicating,lan2022negative}; the appeal to sensation is also common in affective visualization design~\cite{lan2023affective}), it also exhibit distinct features.
First, artistic data visualization can be fundamentally shaped by high-level design paradigms. These paradigms act as the creative lenses that artists use to interpret the world around them, serving as the cornerstone of artistic decision-making. Yet, they are seldom identified in general visualization design research. This also demonstrates artists' emphasis on ideas and concepts.
Second, although artistic data visualization shares some design intents with prior studies, it exhibits more categories and different distributions in frequency of these intents. On one hand, the exploration of novel forms of expression is highly valued by data artists, as evidenced by the highest frequency of \textit{experiment}, a distinctive aspect that sets it apart from other fields. Additionally, within the design intents of artistic data visualization, those that convey strong opinions (\eg \textit{advocate}) are less prevalent. In contrast, higher-ranking intents (\eg \textit{inform}, \textit{engage}, \textit{provoke}) tend to be more implicit, subtle, and thought-provoking. This characteristic aligns well with the traits of contemporary art, which allows the audience significant freedom to interpret the work on their own terms~\cite{pooke2021art}.
Last, regarding specific design techniques, artistic data visualization places a strong emphasis on sensory richness and physicality, with many artworks taking the form of physical installations or artifacts (which is very different from the visualization design genres identified before~\cite{segel2010narrative}). Among current fields, affective visualization design shows the greatest technological affinity with artistic data visualization. However, when examining the distribution and frequency of specific techniques, interactive technologies in artistic data visualization are more prominent, manifested through various body, face, and gesture-based interactions, as well as biomimetic methods. Meanwhile, the use of physical materials is more diverse and pronounced, incorporating a wider range of materials to encode data, such as metals, furs, food, and even microorganisms.

%% file: Sections/05-interview.tex
\section{Perspectives from Data Artists}
\label{sec:interview}

Next, to better understand the underlying considerations behind artistic data visualization, as well as to gain firsthand insights to cross-validate our previous findings, we conducted in-depth interviews with twelve data artists.

\subsection{Participants and Process}
\label{sssec:interviewprocess}

We invited participants through: (i) sending interview invitations to the authors of IEEE VISAP projects, and (ii) reaching out to practitioners who self-tagged as data artists on social media.
A total of 12 data artists accepted our invitation, including 5 females and 7 males. Their ages ranged from 23 to 42 and were diverse in job and educational backgrounds (see \autoref{tab:participants}). 


The interviews were semi-structured. We prepared a set of questions in advance, which can be categorized into four parts: (i) the creation of artistic data visualizations (\eg ``How did you come up with the idea of this project?'', ``How was your design process?''), (ii) the understanding of artistic data visualization (\eg ``What do you think is the most prominent feature of artistic data visualization?'', ``What distinguishes artistic data visualization from other types of data visualization''), (iii) the response to potential critiques (\eg ``If someone challenges the accuracy or efficiency of your visualization, how would you respond?''), (iv) challenges and expectations (\eg ``Have you ever met any challenges?'', ``How do you envision the future of this field?''). 
Before the interviews, we conducted background research on the participants, including their education, key works, achievements, and publications, to ensure the interviews were meaningful and relevant to their experiences and expertise.
During the interviews, we first asked the participants to introduce themselves, as well as their experience with artistic data visualization briefly as a warm-up. Next, we asked the aforementioned interview questions surrounding their representative data artworks. Depending on their responses, we asked follow-up questions to dig deeper into interesting points brought up by them (\eg ``Could you please elaborate on that point?'', ``Can you provide an example to illustrate?''). Each interview lasted about one hour and the interview process was audio recorded with the participant's consent.

To analyze the data, we followed the research methodology suggested by thematic analysis~\cite{braun2022thematic}.
Two authors first read through the transcriptions independently to familiarize ourselves with the data and took notes on initial observations.
Then, we coded the transcriptions with the goal of identifying the responses to the four aforementioned research questions. 
Next, we grouped related codes together to form themes, cross-checked each other's codes and themes, and marked disagreed codes until reaching 100\% consensus. 

\begin{table}[t!]
\fontsize{6.8pt}{7.5pt}\selectfont
\centering
\caption{Information of the interviewees. }
\label{tab:participants}
\vspace{-1em}
\begin{tabularx}{\columnwidth}{p{0.3em}p{0.3em}p{0.5em}p{10.3em}X}
\toprule
ID &	Sex &	Age &	Job &	Educational Background \\
\midrule
P1	&	M	&	26	&	Artist	&	Visual Design	\\
P2	&	M	&	26	&	Product Manager	&	Industrial Design \& Data Visualization \\
P3	&	M	&	39	&	Creative Technologist	& Computer Science \& Visual Design	\\
P4	&	M	&	41	&	Art Studio Head	&	Fine Arts	\\
P5	&	M	&	28	&	PhD Student	&	Architecture \& Computer Science	\\
P6	&	F	&	28	&	Postdoctoral Researcher	&	Computer Science	\\
P7	&	F	&	28	&	UX Designer	&	Digital Media	\\
P8	&	M	&	24	&	Graduate Student 	& Digital Media	\\
P9	&	F	&	43	&	Associate Professor in Art	&	Interior Design \& Multimedia	\\
P10	&	F	&	33	&	Assistant Professor in Art	& Communication Design \\
P11	&	M	&	29	&	Graduate Student	&	Industrial Design	\\
P12	&	F	&	36	&	Artist	&	Media Art	\\
\bottomrule
\end{tabularx}
\vspace{-1em}
\end{table}


\subsection{Findings}

\subsubsection{Creation of Artistic Data Visualization}
We found that artists exhibit some common patterns when ideating and making artistic data visualizations.

\textbf{Motivation.}
We identified four modes of motivation for creating artistic data visualization. 
The first mode is \textbf{value-driven} (P3, P5, P9, P10, P12), which is often based on the artist's high-level values or deep thoughts. For instance, P3 mentioned that his art project originated from his philosophical contemplation of the relationship between humans and nature: ``\textit{A pivotal moment occurred during my journey from Washington to New York, when I was inspired by the mountains outside the window. In contrast, the cityscape and industrial facilities we usually live with are very modern and cold. So, I wondered if I could create a poetic geographical space visualization.}''
The second mode is \textbf{interest-driven} (P6, P8, P10, P11*2 times (corresponding to 2 different works)), which is more personal and related to specific living experiences. For example, P6 is a visualization researcher who has had an interest in art and literature since childhood (``\textit{I have always had a strong interest in the humanities}''). Therefore, after entering the field of visualization, she has always wanted to use visualization techniques to convey the rhythms and emotions found in literature.
The third mode is \textbf{reality-driven} (P1, P7, P9, P11). P7's artwork is about preserving Tibetan calligraphy. She said, ``\textit{We visited the inheritors of Tibetan calligraphy and learned about its long and glorious history. However, very few people are currently aware of this heritage or involved in its preservation. Therefore, we decided to present the unique beauty of this cultural heritage in a visual form.}''
The last mode is \textbf{client-driven} (P1, P2, P4). P4 is a data artist who has his own art studio. His recent project, which involves animating and visually representing ocean tide changes using 3D particle visualization, is based on the specific requirements of a collaborator. Similarly, P2 also mentioned that his art project was a ``\textit{commissioned work}'', meaning that the initial idea had already been determined by someone else.

\textbf{Workflow.}
The workflow of creating artistic data visualizations mainly falls into two categories: input-driven and output-driven. The \textbf{input-driven} workflow is more similar to the classic model of information visualization, which follows an input-output pipeline. 
Artists first obtain a dataset, then analyze it, and create visualizations. Artists often do not know in advance what the data will ultimately lead to. For instance, when working on an art visualization project related to social media, P5 collected data from Twitter after having a preliminary idea, and then tested various data dimensionality reduction algorithms and visualization layouts, finally selecting the most satisfying version.
In contrast, in the \textbf{output-driven} workflow, artists first conceive a mental image of the desired outcome, such as the overall aesthetic style and how the visualization will look like, and then utilize data as a medium to realize it. For instance, P3 had already determined to re-present modern maps using a poetic, painting style before actually beginning to collect and process data. Similarly, when working on her data sonification project, P10 drew inspiration from Kandinsky, who expressed music through points, lines, and planes. She decided to also visualize sound data as geometric shapes before proceeding to collect the necessary data. In terms of frequency, the output-driven workflow (P1, P3, P4, P7, P9, P10, P11*3 times (3 different works), P12) was more common than the input-driven workflow (P1, P2, P5, P6, P8). This finding also resonates with previous research~\cite{tandon2023visual} that artists are especially adept at visual tasks, when compared to mathematicians or computer scientists.
However, despite differences in the overall workflow, all the artists we interviewed mentioned that they iteratively adjust and refine their work until it reaches a satisfactory point. 

\subsubsection{Understanding of Artistic Data Visualization}
\label{sssec:understanding}

Next, we performed an analysis of the participants' descriptions regarding their understanding of artistic data visualization. To delve deeper into this important issue, we employed a richer array of analysis methods. First, we extracted relevant sentences from the participants, conducted word segmentation, synonyms detection, and frequency statistics to identify high-frequency keywords. When analyzing these keywords, close reading was also utilized to ensure a detailed examination of their original contexts.
As a result, the most mentioned words include \textit{express/convey/communicate} (N = 24), \textit{emotion/feeling/subjectivity} (N = 24), \textit{story/narrative} (N = 11), \textit{reflection/critical thinking} (N = 10), \textit{concept/idea} (N = 6), \textit{purpose/intent} (N = 4).
For example, P2 thought artists ``\textit{try to make inherently emotionless data convey emotional effect.}'' P8 said, ``\textit{I believe good data art should be engaging, offering more feelings or allowing people to see things from different perspectives, rather than merely enhancing efficiency.}''
P4 believed that artistic data visualization ``\textit{is not meant to provide a clear answer or a specific analysis of the data; art often involves speculation and reflection.}''
There were also keywords, although mentioned less frequently, that conveyed very interesting values and attitudes, such as \textit{freedom} (N = 2), \textit{openness} (N = 2), and \textit{ambiguity} (N = 2).
For example, P6 said, ``\textit{Art tends to be more divergent and doesn’t have a strong, specific purpose; it requires some ambiguity. We just place it there and let people interpret it however they wish.}'' P12 expressed her view on art with a highly concise statement: ``\textit{Aesthetics is freedom.}'' In her view, artists always strive to break existing rules or constraints, seeking and defending the space for free expression.

These keywords along with their original expressions have some obvious common characteristics.
First, they all involve a \textbf{strong emphasis on human agency}. 
In contrast to science, which prioritizes objectivity, generalizability, and precision~\cite{brown2001art}, artists are more inclined to embrace subjectivity, individuality, and ambiguity.
Second, we noticed that the way artists described artistic data visualization does not focus on specific forms, nor does it emphasize classic aesthetic standards such as harmony or vividness. Instead, they all emphasized the artist's concepts and expressions, demonstrating a \textbf{strong contemporary art characteristic}. 
As introduced in \autoref{ssec:contemporary}, in contemporary times, ``art as imitation'' has been replaced by ``art as expression'' and ``art as concept/idea''~\cite{pooke2021art}. Art is no longer confined to form but is the external manifestation of human concepts and thoughts. 
In general, the perspectives offered by the data artists in our interviews resonate well with the design taxonomy in \autoref{fig:techniques}, and also provide us with a deeper understanding of the underlying logic of the identified techniques (\ie sensation, interaction, narrative, physicality). For example, the emphasis on subjective experience enhances artists' interest in stimulating human senses. The need to communicate ideas and concepts makes them adept at using narratives and eager to experiment with new modes of expression, stimulating the exploration of interactive technologies and various physical materials.

\subsubsection{Response to Critiques or Doubts}
As introduced in \autoref{ssec:artisticvis}, the controversy surrounding artistic data visualization mainly concerns its readability and reading efficiency. In our interviews, while most participants acknowledged that such critiques pose a challenge for artistic data visualization, they also provided some justifications.


\textbf{Artistic data visualization adopts an alternative mechanism of communication.}
5 participants thought that artistic data visualization achieves communicative goals through engaging viewers on a deeper level. For example, P4 argued for a broader understanding of efficiency: ``\textit{If our goal is to see precisely what a data point's value is, artistic data visualization is indeed inefficient...but in reality, the audience is probably not professional data analysts and they don't care about data precision. Instead, they care about what the data conveys and expresses. In that sense, artistic data visualization is efficient. For instance, through audiovisual elements, I can quickly immerse you in a context and convey a message.}''
P5 thought artistic data visualization and conventional statistical charts each have their merits: ``\textit{It's just that the way they convey information is different. With statistical charts, you need to decipher the data and get insights, almost like a bottom-up approach. However, artistic data visualization can directly provide you with messages and insights. If you feel attracted, you can then explore the data within. It's more of a top-down approach.}''
P6 thought there's no need to pursue a universal mode of data communication because: ``\textit{Some people are more rational and some are emotional. Everyone has different tastes. If we can attract those who are willing to understand us and able to understand us, that's already nice.}''

\textbf{Clarity and efficiency are not necessarily sacrificed.}
7 participants stated that their emphasis on subjectivity and humanity does not mean they \textit{only} care about these aspects. Most participants recognized the importance of data precision, particularly at the stages of data collection, preprocessing, and analysis. For example, P5 said, ``\textit{We invested a considerable amount of effort in data scraping and cleaning, just like the general process in data science. You need to address missing values, outliers, and ensure the data is correct.}'' P4 thought that ``\textit{data art should be supported by data; if the data is fictitious, it shouldn't be counted as data visualization.}'' 
Regarding understandability, P1 said, ``\textit{I would carefully consider how users will read and interact with my visualizations. I don't want to create a beautiful non-sense.}''
P3 mentioned that when designing installations for museums, his team would conduct user interviews in advance or perform A/B tests, and ``\textit{if many users failed to understand, we would make design adjustments.}''
P4 thought there were both good and bad designs in artistic data visualization: ``\textit{If a work is completely unreadable, it might indicate poor design. Conversely, if designers can provide readers with legends and reading guidance as much as they can while maintaining aesthetics and novelty, artistic data visualizations can of course be readable.}''
As both a data artist and visualization researcher, P6 said, ``\textit{I do consider users, but any visualization involves trade-offs. Even when creating visual analytics systems, sometimes you sacrifice some utility for innovative visual effects...I don't think artistic data visualization is that special.}''

\subsubsection{Challenges and Expectations}
\label{sssec:expectations}
Lastly, participants reflected on the challenges and their expectations.

\textbf{Facing a bottleneck in skills or creativity.}
First of all, 7 participants mentioned that they faced a bottleneck in issues such as handling big data, crafting a narrative, and thinking of novel ideas. For example, P7 thought ``\textit{conceptualizing a project is challenging}'', and ``\textit{another major challenge is in storytelling...crafting a cohesive narrative is tough.}''
P4 reflected that he would often ``\textit{draw inspiration from the design ideas of outstanding peers, but original and highly innovative ideas totally came up with by myself are not that many.}'' 
P10 said, ``\textit{We currently don’t have access to particularly professional data, and we definitely lack the tools to work with such data.}''
P1 reflected that ``\textit{at present, my work is still largely targeted at the general public and remains a bit shallow. I try to dig deeper...However, if I want to pursue this, I definitely need to explore the deeper scientific insights behind the data.}''

\textbf{Insufficiency of interdisciplinary collaboration \& understanding.}
Relating to the previous issue, 6 interviewees expressed the desire to foster more cross-disciplinary collaborations, such as the hope for scientists to join and enhance professional knowledge of data (P1), and to collaborate with teams specialized in data processing (P10). 
P4 expressed his urgent need to recruit people with expertise in algorithms for his art studio, because ``\textit{to create generative art, such as large-scale data-driven particle systems, one must have knowledge of statistics or machine learning; I have long wanted to recruit individuals with such expertise for my team.}''
However, P6 thought that currently, there is still a lack of mutual understanding among people from different disciplines. She observed that those currently engaged in artistic data visualization hail from diverse backgrounds: some have a technical background and an interest in art, while others are artists by original training: ``\textit{These two groups approach things very differently because their foundational training and theoretical frameworks are completely distinct. Sometimes, even though we claim to be interdisciplinary, I feel like there’s a bias or a gap in understanding between these two groups, as if we’re not speaking the same language.}''

\textbf{The gap between art practice and research.}
Despite the flourish of artistic data visualization in the wild, 6 participants saw a gap between its practice and research. P9, who is both an artist and an assistant professor, highlighted the difficulties of publishing papers and being involved in academia: ``\textit{In art schools, creating works and receiving awards hold more value than publishing papers. Many artists don’t feel the need to publish papers, but due to the larger context of academic evaluation, some are now being pushed to transform their works into papers, which forces them to find new ways of approaching this.}'' P12, who used to be an artist and is now pursuing a Ph.D. degree, said that ``\textit{I have created many works and participated in many exhibitions, but I truly don't know how to write a doctoral dissertation about them...It feels like the thought processes for conducting research and creating artwork are quite different.}''
P8 felt that ``\textit{it (the academic community) offers limited theoretical support for my work.}''
P4 recalled having moments of self-doubt because ``\textit{we produce a lot of work, but the methodologies and frameworks for organizing this kind of work are scarce...sometimes it's hard to justify our value to others.}''

%% file: Sections/06-discussion.tex
\section{Lessons for Future Work}
\label{sec:discuss}

In this work, we adopt a multi-faceted approach to explore artistic data visualization. Our literature review, corpus analysis of artistic data visualizations, and interviews with data artists collectively indicate that \textbf{artistic data visualization is deeply rooted in art discourse, with its aesthetic pursuits particularly contextualized within contemporary art philosophies and values. This fundamentally distinguishes artistic data visualization from data science in terms of epistemological foundations and methodological approaches to interpreting and presenting information.} 
Thus, we are prompted to ask: How can data art add value to the visualization community? What are the areas where artists and scientists can learn from and inspire each other?
Below, we organize the lessons we've learned (marked as L1-L7) in three subsections, corresponding to three areas of visualization research. 

\subsection{Theoretical \& Empirical Work}

Firstly, art and science share many dialogic points in concept establishment and problem-solving methods.

\textbf{L1: Broadening the understanding of aesthetics.}
Previously, the visualization community has often operationalized aesthetics as beauty and hedonic features, and specific scales (\eg enjoyable, likable, pleasing) have been developed to measure aesthetic pleasure~\cite{he2022beauvis,stoll2024investigating}. 
However, taking artistic data visualization as an example, aesthetics can extend far beyond sensory beauty to encompass intellectual levels of critical, deconstructive, or rebellious stances towards established norms. 
Besides, as summarized in \autoref{ssec:contemporary}, technological innovation has consistently been a significant driver of aesthetic evolution. Just as the invention of photography pushed art beyond mimesis, the rise of AI has led to challenges such as copyright infringement, economic loss, and the undermining of art's uniqueness and originality~\cite{jiang2023ai}. Such challenges may further compel artists to pursue newer, more rebellious aesthetics. This shift is already being recognized in the art world with terms such as \textit{computational aesthetics}~\cite{galanter2012computational}, and in this work, we have seen artworks that embrace glitches and errors in machines, an appreciation for the randomness inherent in machine learning (\eg \autoref{ssection:machine}). Some works even celebrate anti-beauty: \autoref{fig:techniques} (L) may appear terrifying and disturbing to some viewers, but it is intentionally crafted to visualize the world through the lens of AI models, prompting reflections on our increasingly machine-driven society. In a word, future aesthetics are likely to reflect a broader range of creative expressions and critical perspectives, and simply operationalizing aesthetics as beauty and hedonic features may be insufficient. A deeper and more diverse understanding of aesthetics is needed.



\textbf{L2: Enriching evaluation metrics.} 
Following L1, an emerging issue is how to capture the richness and profundity of aesthetics. In more practical terms, we still need to study and explain why certain data artworks are so successful and captivating. Regarding this issue, Shusterman's ~\cite{shusterman2000pragmatist} reflection on \textit{analytical aesthetics} may be helpful. Shusterman argued that \textit{analytical aesthetics}, which treats aesthetics as a set of objective and low-level attributes that can be measured independently, is insufficient. The primary shortcoming of this approach is that it overlooks the complex context of aesthetic experience. Therefore, he called for measuring \textit{pragmatist aesthetics} to better capture the contextual and subjective experience of art. In the context of artistic data visualization, this can include considering viewers' environment when encountering and interacting with data artworks, how the artworks evoke their existing knowledge and values, and their mental journeys. Moreover, as revealed by this work, contrasting with other information displays designed for readability and efficiency, art often enables open-ended interpretation and repeated contemplation. Therefore, it is also crucial to assess long-term user experience beyond capturing immediate responses. To uncover these issues, we view methods such as in-depth interviews, user diaries, longitudinal surveys, and ethnography as valuable complements to traditional research methods.

\textbf{L3: Incorporating cross-disciplinary methods.}
Data artists have their own ways of problem-solving. Many artists have already explored using data visualization to actively address real-world challenges. Such practices align well with research work that seeks novel approaches and innovative visual solutions to solve domain-specific problems.
First, during the ideation stage, the creative methods employed by artists, such as brainstorming, hackathons, and workshops, may be worth learning from. Kerzner~\etal~\cite{kerzner2018framework}, for example, constructed a set of guidelines for creative visualization-opportunities workshops (CVO) to speed up the early stages of applied visualization design by adapting workshop methods from other disciplines.
Then, in the design phase of visualization, researchers can learn from artists' design approaches, exploring more genres, modes of expression, and user engagement methods. For instance, guided by the idea of participatory design, Perovich~\etal~\cite{perovich2020chemicals} visualized river pollution data as river lanterns and organized an event to engage residents in collective reflection on the environmental issue.
Next, in the deployment phase, the methods artists use to implement visualizations are also inspiring. These include collaborating with communities and stakeholders, as well as deploying visualizations in various offline scenarios and everyday settings (\eg ~\cite{rodgers2011exploring,offenhuber2019data}).
Conversely, artists can also benefit from established methodologies in academia. For example, to bridge the gap between practice and research identified in \autoref{sssec:expectations}, they might adopt the design study methodology from visualization research~\cite{sedlmair2012design}. This approach can help artists clearly define the problems they encounter in their creative process. From ideation to final realization, each step can be better justified, providing a more structured approach to complement their creative instincts. Additionally, they might learn from the criteria for rigor (\eg reflexive, plausible) proposed by visualization researchers~\cite{meyer2019criteria} to create more reliable data-driven works.

\subsection{Representation \& Interaction}

Secondly, artists and scientists may inform each other when pushing the boundaries of visual and interactive techniques.

\textbf{L4: Learning from artistic design schemes.} 
Borrowing or transferring existing design schemes from artworks to visualization design is an interesting way to explore novel design. 
On the one hand, researchers can consider transferring design schemes across different dimensions (\eg color, texture, metaphor) and granularities (\eg atomic element, overall style). For example, Samsel~\etal~\cite{samsel2018art} focused on extracting color palettes from paintings to encode scientific data. By contrast, Kyprianidis~\etal~\cite{kyprianidis2012state} implemented algorithms to transfer entire artistic painting styles (\eg Impressionism, cartoon styles) to images and videos.
On the other hand, since these design schemes will ultimately be applied to data, the compatibility of the transferred design with the data should be considered. For instance, taking inspiration from Mondrian's abstract art, Holmquist et al.~\cite{skog2003between,holmquist2003informative,redstrom2000informative} redesigned visualizations such as weather forecasts and bus departure information; during the design process, they conducted multiple iterations to ensure readability and understandability. 
When transferring emotional patterns in motion graphics (\eg bounce, wiggle) to the design of animated charts, Lan et al.~\cite{lan2021kineticharts} conducted multiple rounds of studies to modify distracting animation designs. 
Therefore, while transferring artistic schemes to visualizations is innovative, we recommend conducting rigorous testing and evaluations to confirm its validity.

\textbf{L5: Exploring alternative encoding channels.}
In this work, we see data artists exploring sound, touch, smell, and taste, and experimenting with various materials to encode data. These practices resonate well with the trend of expanding data encoding modalities and extending interactions beyond desktops and smartphones in the visualization community. For instance, many artworks have provided vivid and pioneering examples for data physicalization~\cite{hornecker2023design} and sonification~\cite{enge2024open}, enabling people to engage with data visualization across a broader range of channels and daily contexts.
Such exploration is also relevant to research on accessible visualization, as it facilitates the study of replacement strategies for visual encodings (\eg how to present visualization to visually impaired individuals). Although data sonification is an emerging approach~\cite{holloway2022infosonics}, designing effective sonification remains challenging. Concurrently, data artists possess extensive experience in designing cross-modal visualizations, which could potentially offer valuable insights into making visualizations more accessible to a broader audience.


\subsection{Applications}

Lastly, for application development, we propose that such systems can be developed either \textit{for} or \textit{with} data artists.

\textbf{L6: Developing tools \textit{for} data artists or users who want to design like artists.}
Tools can be developed to serve the needs and lower the barrier of creating artistic data visualizations. For example, Schroeder~\etal~\cite{schroeder2015visualization} presented an interface to assist artists in creating visualizations through direct painting or sketching on a digital data canvas. Inspired by the data art project, \textit{Dear Data} (a year-long postcard exchange of hand-drawn data visualizations, showcasing the two artists' personal data), tools such as Dataselfie~\cite{kim2019dataselfie} were developed to empower users to easily create their own personal visualizations in hand-drawn or pictorial styles.
Additionally, to spark creativity and facilitate ideation, tools can be developed to help users explore, craft, and play with various visualization choices. DataQuilt~\cite{zhang2020dataquilt}, for example, is an interactive tool that allows users to extract elements from raster images and encode data to them, exploring various possibilities of pictogram design.

\textbf{L7: Developing systems \textit{with} data artists.}
In line with prior studies~\cite{judelman2004aesthetics,samsel2013art,li2021we,sandin2006artist}, we found that data artists and scientists/engineers share common pursuits such as exploring new problems and finding appropriate ways to express information; also, a significant number of data artists are highly skilled in programming and interested in software development.
Building on these commonalities, the involvement of artists is expected to offer additional perspectives and knowledge for system development. For example, artists' expertise in manipulating visuals and user studies~\cite{tandon2023visual,sandin2006artist} may help enhance the expressiveness and usability of a system.
Furthermore, as found in both our corpus analysis and interviews, data artists excel at conceptualizing values and meanings, often needing a clear \textit{why} before addressing the \textit{how}. This likely explains why artists were found to be effective at challenging conventional views and offering insights into fundamental questions, such as the necessity and purpose of a system~\cite{laidlaw1998art,pousman2007casual}.
As concluded by Gates~\cite{gates2016art}, ``One extremely valuable role that artists can play in science and with scientists is to ask questions. Even though science is at heart a questioning process...it is easy to lose sight of some of the basic questions that motivate more detailed and specific research.'' 

\section{Limitations}

Our analysis of artistic data visualizations is based on a self-constructed corpus. While we endeavored to include a diverse range of sources, our collection primarily features works from or closely associated with the visualization academic community. 
Also, since we required artworks or their publication venues to explicitly mention keywords such as artistic data visualization and data art, some works may not have been included in our study due to a lack of description or tagging.
Thus, it is not exhaustive and does not represent all artistic data visualizations in the wild. Also, it contains a relatively small number of early data artworks due to issues such as lack of digitization or broken links, as well as a scarcity of non-Western artworks. 
In addition, due to the nature of the interview research method, the number of data artists we were able to interview is limited. 
We hope to engage with more practitioners in the field of artistic data visualization to obtain more first-hand and in-depth research materials. 
Besides, when performing thematic analyses, while we achieved consensus among coders, some nuances might have been sacrificed. For instance, the concept of \textit{sculpture}, which has been frequently used by artists, was initially coded but later dropped due to its broad applicability, ranging from static physical objects to various interactive installations. This decision, aimed at reducing ambiguity, may have led to a loss of deeper connotations, such as the heritage relationships between data art and traditional fine arts. This also highlights the need to enhance the reflectiveness of coders~\cite{braun2022thematic} when conducting qualitative coding.

%% file: Sections/07-conclusion.tex
\section{Conclusion}

In this work, we have analyzed a corpus of data artworks and conducted interviews with data artists to understand the design features and practical perspectives of artistic data visualization, as well as to draw its implications for the visualization community. In brief, our findings indicate that artistic data visualization has inherited many characteristics from the field of art, which gives it distinct features in design concepts, methods, and processes compared to visualizations centered around analytics. These characteristics can, on one hand, deepen our understanding of how artists are pushing the boundaries of visualization applications. On the other hand, the inspiring thoughts and practices of data artists can also bring new perspectives and opportunities for future visualization research.